\documentclass[11pt,letterpaper]{article}
\pdfoutput=1
\usepackage{jcappub}
\usepackage{bbm}
\usepackage{mathrsfs}
\usepackage{slashed}
\usepackage{caption}
\usepackage{epstopdf}
\usepackage[normalem]{ulem}
\usepackage[bottom]{footmisc}
\usepackage{subcaption}
\usepackage{bbold}
\usepackage{titlesec}
\usepackage{threeparttable}
\usepackage{booktabs}
\usepackage{changepage}
\usepackage[utf8]{inputenc}
\usepackage{dsfont} 
\usepackage{grffile}
\usepackage{graphicx}  
\usepackage{dcolumn}   
\usepackage{bm}        
\usepackage{amssymb}   
\usepackage{setspace}
\usepackage{amsmath, amssymb, setspace}
\usepackage{array}
\usepackage{booktabs}
\usepackage{caption}
\usepackage{float}
\usepackage{lmodern}
\usepackage{multirow}
\usepackage{soul}
\usepackage[normalem]{ulem}
\usepackage{braket}
\usepackage{comment}
\usepackage[draft]{pgf}
\usepackage{adjustbox} 
\usepackage{xspace} 
\usepackage{url}
\usepackage{xcolor}
\usepackage{comment}

\usepackage{orcidlink}

\usepackage{indentfirst}
\newcommand{\mi}{\mathrm{i}}
\newcommand{\me}{\mathrm{e}}

\newcommand{\tr}{\mathrm{Tr}}
\newcommand{\lp}{\left(}
\newcommand{\rp}{\right)}
\newcommand{\ls}{\left[}
\newcommand{\rs}{\right]}

\allowdisplaybreaks

\newcommand{\be}{\begin{equation}}
\newcommand{\ee}{\end{equation}}
\def \d {{\rm d}}

\newcommand{\bea}{\begin{eqnarray}}
\newcommand{\eea}{\end{eqnarray}}

\title{Identifying the Quadrupolar Nature of Gravitational Wave Background through Space-based Missions}

\author{
Yifan Chen\,\orcidlink{0000-0002-2507-8272}$^{a}$\footnote{Corresponding author.},
Yuxiang Liu\,\orcidlink{0009-0000-1749-8149}$^{b}$,
Jing Shu\,\orcidlink{0000-0001-6569-403X}$^{b,c,d}$\footnote{Corresponding author.},
Bin Xu\,\orcidlink{0000-0003-4881-9229}$^{b}$,
Xiao Xue\,\orcidlink{0000-0002-0740-1283}$^{e}$, 
and Yanjie Zeng\,\orcidlink{0000-0002-5175-7797}$^{f,g}$
}

\affiliation{
$^a$Center of Gravity, Niels Bohr Institute, Blegdamsvej 17, 2100 Copenhagen, Denmark\\
$^b$School of Physics and State Key Laboratory of Nuclear Physics and Technology, Peking University, Beijing 100871, China\\
$^c$Center for High Energy Physics, Peking University, Beijing 100871, China\\
$^d$Beijing Laser Acceleration Innovation Center, Huairou, Beijing, 101400, China\\
$^e$Institut de Física d’Altes Energies (IFAE), The Barcelona Institute of Science and Technology, Campus UAB, 08193 Bellaterra (Barcelona), Spain\\
$^f$Institute of Theoretical Physics, Chinese Academy of Sciences, Beijing 100190, China\\
$^g$School of Physical Sciences, University of Chinese Academy of Sciences, Beijing 100049, China
}

\emailAdd{yifan.chen@nanograv.org}
\emailAdd{existent@pku.edu.cn}
\emailAdd{jshu@pku.edu.cn}
\emailAdd{binxu@kias.re.kr}
\emailAdd{xxue@ifae.es}
\emailAdd{zengyanjie@itp.ac.cn}

\abstract{The stochastic gravitational wave background (SGWB) consists of an incoherent collection of waves from both astrophysical and cosmological sources. To distinguish the SGWB from noise, it is essential to verify its quadrupolar nature, exemplified by the cross-correlations among pairs of pulsars within a pulsar timing array, commonly referred to as the Hellings-Downs curve. We study how this quadrupolar signature manifests in correlations between general GW detector pairs, characterized by their antenna responses. Focusing on space-based missions—including laser interferometers (LISA, Taiji, TianQin) and atom interferometers (AEDGE/MAGIS)—we demonstrate how orbital motion dynamically modulates detector correlations through time-dependent separations and relative orientation shifts. These modulations encode unique statistical features that serve as definitive markers of the SGWB’s quadrupolar nature. Our findings identify optimal configurations for these missions, offer forecasts for the time needed to identify the quadrupolar nature of the SGWB, and are applicable to both space-space and space-terrestrial correlations.}

\begin{document}
\maketitle
\flushbottom

\section{Introduction}

The first observation of gravitational waves (GWs) from a black hole (BH) merger by LIGO and Virgo~\cite{LIGOScientific:2016aoc} has opened a new avenue for probing gravity, astrophysics, and cosmology. More recently, multiple pulsar timing array (PTA) collaborations have reported the discovery of a stochastic GW background (SGWB)~\cite{NANOGrav:2023gor,EPTA:2023fyk,Reardon:2023gzh,Xu:2023wog}. This significant milestone was initially identified as a common-spectrum process~\cite{Arzoumanian:2020vkk} and subsequently confirmed through the correlations among different pulsars, known as the Hellings-Downs curve~\cite{Hellings:1983fr}. The Hellings-Downs curve is a pivotal characteristic of the quadrupolar tensor nature of the GWs detected by PTAs. The amplitude and spectral index of the inferred SGWB are consistent with those predicted from a population of supermassive BH binaries (SMBHB)~\cite{NANOGrav:2023hfp,EPTA:2023xxk,NANOGrav:2024nmo}.

As we enter the golden age of GW detection, new experiments and proposals continue to emerge, building on the successes of terrestrial laser interferometers and PTAs. These include space-based laser interferometers~\cite{LISA:2017pwj,TianQin:2015yph,Ruan:2018tsw}, atom interferometers~\cite{Dimopoulos:2007cj,Graham:2017pmn,Badurina:2019hst,AEDGE:2019nxb}, astrometric measurements~\cite{Braginsky:1989pv,Book:2010pf,Moore:2017ity,Garcia-Bellido:2021zgu,Wang:2022sxn,Jaraba:2023djs,Caliskan:2023cqm,An:2024axz}, binary resonance~\cite{Hui:2012yp,Blas:2021mpc,Blas:2021mqw}, and various tabletop experiments for high-frequency GWs~\cite{Aggarwal:2020olq} These diverse detectors respond differently to GWs, characterized by their unique antenna patterns~\cite{Romano:2016dpx}.

To distinguish an SGWB from noise, correlations between different detectors are essential to reveal the quadrupolar nature of the signals, similar to how PTAs resolve the Hellings-Downs curve. Correlations between a pair of detectors are referred to as overlap reduction functions (ORFs)~\cite{Finn:2008vh}, or generalized Hellings-Downs correlations, which depend on the distance and relative orientations of the detectors. Furthermore, the ORFs for observing an SGWB are characterized by both their mean value after ensemble averages and their variance due to the stochastic nature of the GWs~\cite{Allen:2022dzg,Allen:2022ksj,Bernardo:2022xzl,Bernardo:2023bqx,Bernardo:2023pwt,Romano:2023zhb,Bernardo:2023zna,Caliskan:2023cqm,Agarwal:2024hlj,Bernardo:2024bdc}.

In this work, we investigate the identification of the quadrupolar signature in SGWBs using general GW detector networks. We present case studies involving space-based missions: correlations between the solar-orbiting laser interferometers LISA~\cite{LISA:2017pwj} and Taiji~\cite{Ruan:2018tsw}, and between the Earth-orbiting laser interferometer TianQin~\cite{TianQin:2015yph} and the atom interferometers AEDGE~\cite{AEDGE:2019nxb}/MAGIS~\cite{Graham:2017pmn}. Our framework extends straightforward to space-terrestrial correlations, such as those combining the Einstein Telescope~\cite{Punturo:2010zz}, Cosmic Explorer~\cite{Evans:2021gyd}, and AION~\cite{Badurina:2019hst}. Crucially, the orbital motion of space-based detectors enables continuous scanning of orientation angles, dynamically probing the ORF across its parameter space. Leveraging this capability, we propose optimized mission configurations and forecast the time required to resolve the quadrupolar nature of an SGWB, relative to the duration needed for its initial detection as a power excess above noise.

Our framework is general and applicable to a wide range of SGWBs, including astrophysical sources such as stellar-mass compact binaries~\cite{LIGOScientific:2016fpe,KAGRA:2021duu,Banks:2023eym} as well as potential cosmological signals. We adopt the stellar binary background as a benchmark for quantitatively estimating the timescale required to identify the quadrupolar nature, as this background is irreducible and its spectral amplitude and index are currently subject to relatively small astrophysical uncertainties. The framework can also be readily extended to cosmological sources, which may produce signals with amplitudes significantly larger than those of the astrophysical foreground. To accommodate this broader class of sources, we develop formalisms applicable to both the {\emph{weak}} and {\emph{strong signal}} regimes.

The structure of the paper is as follows. In Sec.~\ref{sec:SGWBC}, we review the fundamentals of the SGWB and explain how its quadrupolar nature manifests in the cross-correlations of detector pairs. Section~\ref{sec:statistics} presents an analytical statistical framework for evaluating the duration required to identify the quadrupolar nature of the SGWB, in comparison to the initial detection of a power excess, under both the \emph{weak} and \emph{strong signal limits}. In Sec.~\ref{sec:IQNS}, we examine near-future space missions and demonstrate how their self-rotation and orbital motion can be used to scan the correlation space and uncover the quadrupolar signature. Section~\ref{sec:discussion} concludes with a summary and outlook for more general detector networks.

\section{Stochastic Gravitational Wave Detection via Detector Pairs}\label{sec:SGWBC}
This section reviews how correlations between detector pairs can be used to identify the quadrupolar nature of the SGWB. We begin by parameterizing the SGWB, then introduce a generalized antenna pattern basis for GW detectors, and finally compute their correlations.

\subsection{Stochastic Gravitational Wave Background}
An SGWB is an aggregate of GWs from various directions and with random phases, originating either cosmologically or astrophysically. The nanohertz SGWB was recently detected by PTAs, with a spectrum consistent with predictions from the inspiral of SMBHBs~\cite{NANOGrav:2023hfp,EPTA:2023xxk,NANOGrav:2024nmo}. Stellar-mass compact binaries, including BHs and neutron stars, are expected to contribute to a similar background across frequencies ranging from $10^{-4}$~Hz to $10^2$~Hz, detectable by next-generation terrestrial and space laser interferometers, and long-baseline atom interferometers~\cite{LIGOScientific:2016fpe,KAGRA:2021duu,Banks:2023eym}.

Working within the transverse-traceless (TT) gauge for GWs and adopting the speed of light  $c=1$  unit, GWs are modeled as perturbations of the spacetime metric. These perturbations can be represented as a superposition of plane waves emanating from direction $\hat{k}$ with frequency $f$:
\begin{equation}
    h_{ab}(t,\vec{x})=\int_{-\infty}^{+\infty} \d f \int \d^2\hat{k} \, h_{ab}(f,\hat{k}) \, \me^{ \mi 2\pi f(t - \hat{k}\cdot \vec{x})}.
    \label{h_def}
\end{equation}
Here, the Fourier coefficients  $h_{ab}(f,\hat{k})$ represent the strain in the frequency domain and can be expanded based on GW polarizations:
\begin{equation}
    h_{ab}(f,\hat{k})=h_+(f,\hat{k}) \, \hat\epsilon^+_{ab}(\hat{k})+h_{\times}(f,\hat{k}) \, \hat\epsilon^{\times}_{ab}(\hat{k}),
\end{equation}
where $h_+(f,\hat{k})$ and $h_\times(f,\hat{k})$ are 
the respective mode functions. The linear polarization basis tensors are defined as
\begin{equation}
\begin{aligned}
    \hat\epsilon^+_{ab}(\hat{k}) &=\hat{l}_a \hat{l}_b - \hat{m}_a \hat{m}_b,\\
    \hat\epsilon^{\times}_{ab}(\hat{k}) &=\hat{l}_a \hat{m}_b + \hat{m}_a \hat{l}_b,
    \label{GWlp}
\end{aligned}
\end{equation}
where $\hat{l}$ and $\hat{m}$ are orthonormal vectors to $\hat{k}$, defined using the spherical angles $(\theta_k, \phi_k)$ as
\begin{equation}
\begin{aligned}
    \hat{k} & \equiv \left(\sin \theta_k \cos \phi_k,\sin \theta_k \sin \phi_k,\cos \theta_k\right)^{T},\\
    \hat{l} & \equiv \left(\cos \theta_k \cos \phi_k,\cos \theta_k \sin \phi_k,-\sin \theta_k \right)^{T}, \\
     \hat{m} & \equiv \left(-\sin \phi_k,\cos \phi_k,0 \right)^{T}.
\end{aligned}
\end{equation}

The statistical properties of a Gaussian, stationary, unpolarized, and spatially homogeneous and isotropic SGWB are described by the first and second moments of the strain:
\begin{equation}
    \langle h_A(f,\hat{k})\rangle=0, \qquad
    \langle h_A(f,\hat{k}) h^*_{A'}(f',\hat{k}')\rangle=\frac{S_h(f)}{16\pi}\delta(f-f')\delta_{AA'}\delta^2(\hat{k},\hat{k}'),
\end{equation}
where $\langle \cdots \rangle$ denotes the ensemble average over the SGWB, $A$ labels the GW polarization state, and $S_h(f)$ is the one-sided power spectral density (PSD) of the strain. The total energy density in GWs is given by~\cite{Moore:2014lga}:
\begin{equation}
    \rho_{\rm GW}=\int_{0}^{f_{\infty}} \d f \frac{\pi f^2}{4G} S_h(f),
\end{equation}
with $G$ representing the Newtonian constant.

\subsection{Detector Responses}
A GW detector typically responds linearly to GWs, converting them into signals. Representing the time series of channel $I$ by $d_I(t)$, its transformation into the frequency domain is given by:
\begin{equation}
    d_I(f)=\int_{-\infty}^{\infty} \d t \, \me^{-2\pi i ft}d_I(t) ={s}_I(f)+{n}_I(f),
\end{equation}
where $s_I(f)$ and $n_I(f)$ denote the GW signal and measurement noise contributions, respectively. The signal component is formulated as:
\begin{align}
    {s}_I(f)=\int \d^2\hat{k} \, R_I^{ab}(f, \hat{k}) \, h_{ab}(f,\hat{k}) \, \me^{-i2\pi f\hat{k}\cdot \vec{x}}
\end{align}
where $R_I^{ab}(f, \hat{k})$ is the response tensor of the detector located at $\vec{x}$. As the TT-gauge strain $h_{ab}$ is symmetric and traceless, only the symmetric and traceless components of $R_I^{ab}(f, \hat{k})$ contribute to detectable signals.

In the \emph{small-antenna limit}, where the detector size is significantly smaller than the GW wavelength $1/f$, the response tensor becomes independent of $\hat{k}$. The response matrix $R^{ab}_I$ for channel $I$ can be decomposed into a basis of symmetric, traceless polarization tensors as
\begin{equation}
R^{ab}_I = \sum_m B^m_I \, \hat{R}^{ab}_m,
\end{equation}
where $B^m_I$ are channel-dependent response coefficients, and $\hat{R}^{ab}_m$ form a convenient orthonormal basis. A widely used choice is the generalized tensor polarization basis~\cite{Will:1993hxu, Zhou:1995en, Maggiore:2007ulw}:
\begin{align}
	\hat R_{T+}&=(\hat e_x\otimes \hat e_x-\hat e_y\otimes \hat e_y)/\sqrt{2}, & \hat R_{T\times}&=(\hat e_x\otimes \hat e_y+\hat e_y\otimes \hat e_x)/\sqrt{2}, \nonumber\\
	\hat R_{Vx}&=(\hat e_x\otimes \hat e_z+\hat e_z\otimes \hat e_x)/\sqrt{2}, & \hat R_{Vy}&=(\hat e_y\otimes \hat e_z+\hat e_z\otimes \hat e_y))/\sqrt{2},\nonumber\\
	\hat R_{SL}&=(\hat e_z\otimes \hat e_z - \mathbf{I}_3/3)\,\sqrt{3/2}, & &
 \label{eq:RLB}
\end{align}
where $T+, T\times, Vx, Vy,$ and $SL$ denote the tensor-plus, tensor-cross, vector-$x$, vector-$y$, and scalar-longitudinal modes, respectively. These basis tensors satisfy the orthonormality condition $\mathrm{Tr}(\hat R_m \hat R_n^\dagger) = \delta_{mn}$ and align with the standard GW polarization conventions. $\hat{e}_x, \hat{e}_y$ and $\hat{e}_z$ are unit directional vectors of the Cartesian coordinate system, and $\mathbf{I}_3$ represents the $3\times3$ identity matrix.

\begin{figure}[htbp]
    \centering
    \includegraphics[width=0.5\linewidth]{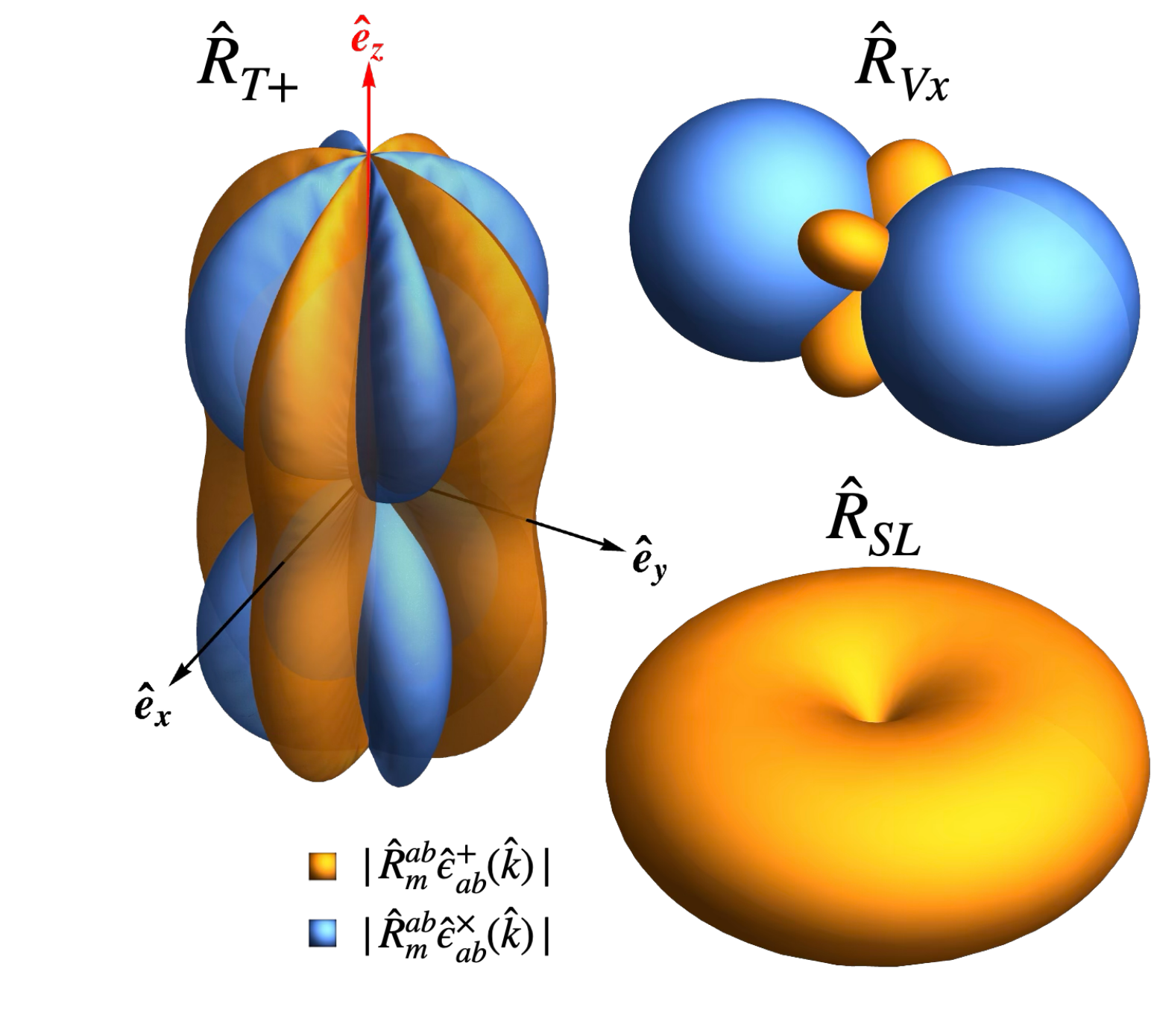}
    \caption{Antenna patterns $|\hat{R}^{ab}_m \hat{\epsilon}_{ab}^A (\hat{k})|$ depicted for three detector bases $m$: $T+$ (left), $Vx$ (top-right), and $SL$ (bottom-right) for GW propagating from the directions $-\hat{k}$, with polarization $A=+$ (orange) and $A=\times$ (blue). The bases $T\times$ and $Vy$ correspond to those of $T+$ and $Vx$, respectively, but are rotated by $\pi/4$ and $\pi/2$ along the $\hat e_z$-axis.}
    \label{fig:antenna}
\end{figure}

In Fig.~\ref{fig:antenna}, we illustrate the antenna patterns $|\hat{R}^{ab}_m \hat{\epsilon}_{ab}^A (\hat{k})|$ for $m=T+$ (left), $Vx$ (top-right), and $SL$ (bottom-right) for a GW propagating in the direction $\hat{k}$ with the polarization tensor $\hat{\epsilon}_{ab}^A$. The orange and blue regions correspond to $A = +$ and $\times$ polarizations, respectively. The antenna patterns for the $T\times$ and $Vy$ bases are identical to those of $T+$ and $Vx$, with rotations of $\pi/4$ and $\pi/2$ along the $\hat e_z$-axis, respectively. 

For instance, a Michelson-type laser interferometer such as LIGO measures the phase difference between two laser beams along perpendicular arms, corresponding to the response matrix $R^{ab}_I = \hat{R}_{T+}$, with arms aligned along $\hat{e}_x$ and $\hat{e}_y$. LISA-like space-based interferometers use both $\hat{R}_{T+}$ and $\hat{R}_{T\times}$ response tensors, with $\hat{e}_z$ perpendicular to the detector constellation plane~\cite{Gair:2012nm}. Long-baseline atom interferometers measure laser pulses propagating along a specific direction, correspond to the $\hat{R}_{SL}$ response matrix, with the $\hat{e}_z$-axis aligned with the laser propagation direction. PTAs similarly have scalar-longitudinal responses, but with an additional redshift factor $B_I^{SL} \propto 1/(1+\hat{k}\cdot \hat e_z)$, aligned with the pulsar line-of-sight~\cite{Anholm:2008wy,Book:2010pf}. 

Linear combinations of the bases above yield the following circular bases $\hat R_m$, with $m$ denoting the azimuthal number:
\begin{equation}
    \hat R_0=\hat R_{SL},\quad \hat R_{\pm1}=\frac{1}{\sqrt{2}}(\mp\hat R_{Vx}-i\hat R_{Vy}),\quad \hat R_{\pm2}=\frac{1}{\sqrt{2}}(\hat R_{T+}\pm i\hat R_{T\times}).
    \label{eq:RCB}
\end{equation}
Here, $\hat R_m$ also satisfies a similar normalization condition: $\mathrm{Tr}(\hat R_{m}\hat R_{n}^\dagger)=\delta_{mn}$.

\subsection{Correlations and Overlap Reduction Functions}
Within a network of detectors, each detector may have multiple response channels, each characterized by a basis as outlined in the previous subsection. The SGWB signal in any single channel is indistinguishable from uncalibrated measurement noise. However, correlations between pairs of channels can effectively distinguish between these two components, as encapsulated by the equation:
\begin{equation}\label{avg}
    \langle d_I(f)d_J^*(f^{\prime}) \rangle=\langle {s}_I(f){s}_J^*(f^{\prime}) \rangle+\langle {n}_I(f){n}_J^*(f^{\prime}) \rangle=\frac{1}{2} \ls S_h(f)\gamma_{IJ}(f)+\delta_{IJ}S_{n_I}(f)\rs \delta(f-f^{\prime}),
\end{equation}
where $I, J$ denote the response channels, with $I, J = 1, 2, \ldots, N_{\rm ch}$ for a total of $N_{\rm ch}$ channels. We assume that the SGWB and the measurement noise are uncorrelated, i.e., $\langle {s}_I(f){n}_J^*(f^{\prime}) \rangle=\langle {n}_I(f){s}_J^*(f^{\prime}) \rangle=0$, and that the noise is stationary, Gaussian, and uncorrelated between different channels:
\begin{equation}
    \langle{n}_I(f){n}_J^*(f')\rangle=\frac{1}{2}\delta(f-f')\delta_{IJ}S_{n_I}(f).
\end{equation}
Here, $S_{n_I}(f)$ is the one-sided measurement noise PSD of the $I$-th channel. 
The ORF, $\gamma_{IJ}(f)$, quantifies the correlation between two detector channels~\cite{Allen:1997ad}. Assuming the SGWB is isotropic and unpolarized, it is given by
\begin{equation}\label{eq:ORFS}
\gamma_{IJ}(f) \equiv \sum_{m_I,m_J} \hat{R}^{ab}_{m_I} \hat{R}^{cd*}_{m_J} \int \frac{\d^2 \hat{k}}{{8\pi}} B_I^{m_I} B^{m_J*}_{J}
    \sum_A \hat{\epsilon}^{A}_{ab}(\hat{k}) \hat{\epsilon}^{A*}_{cd}(\hat{k}) e^{i2\pi f\hat{k}\cdot (\vec{x}_J-\vec{x}_I)},
\end{equation}
where $\hat{\epsilon}^{A}_{ab}$ are the GW polarization tensors.

Additionally, the variance of the correlated data $d_I(f)d_J^*(f)$ can be calculated as:
\begin{equation}
\begin{split}
&\langle |d_I(f)d_J^*(f)|^2\rangle-|\langle d_I(f)d_J^*(f)\rangle|^2
\\=&\frac{T^2}{4}\ls\left(\gamma_{II}(f)S_h(f)+S_{n_I}(f)\right)\left(\gamma_{JJ}(f)S_h(f)+S_{n_J}(f)\right) +\gamma_{IJ}^2(f) S_h^2(f)\rs.
\end{split}
\end{equation}
In this expression, Eq.~\eqref{avg} is applied to derive $\langle d_I(f)d_J^*(f)\rangle=\ls S_h(f)\gamma_{IJ}(f)+\delta_{IJ}S_{n_I}(f)\rs T/2$, where the delta function is replaced by the observation time $T$. Even in the \emph{strong signal limit}, where $S_h \gg S_{n_{I/J}}$, the stochastic nature of GWs induces a significant variance in the correlated data. This quantity, referred to as the total variance, is given by~\cite{Allen:2022dzg,Bernardo:2022xzl}:
\be \Delta\gamma_{IJ} = \left(\gamma_{II} \gamma_{JJ} + \gamma_{IJ}^2 \right)^{1/2}.\ee
Both the expected mean correlation in Eq.~(\ref{eq:ORFS}) and the total variance encode the quadrupolar nature of the SGWB.

Defining the baseline vector $\vec{u} \equiv \vec{x}_J - \vec{x}_I$, with $u = |\vec{u}|$ and $\hat{u} = \vec{u}/u$, and assuming $\hat{k}$-independent coefficients $B_I^m$, the general form of the ORF for symmetric and traceless response matrices can be simplified from Eq.~(\ref{eq:ORFS}) as follows~\cite{Flanagan:1993ix, Allen:1997ad}:
\begin{equation}\label{full}
\gamma_{IJ}=\rho_1(ku)R_I^{ab}R_J^{*ab}+\rho_2(ku)R_I^{ab}R_J^{*ac}\hat{u}^b\hat{u}^c+\rho_3(ku)R_I^{ab}R_J^{*cd}\hat{u}^a\hat{u}^b\hat{u}^c\hat{u}^d,
\end{equation}
where $\rho_1, \rho_2$ and $\rho_3$ are defined as linear combinations of spherical Bessel functions of the first kind $j_n(ku)$~\cite{Allen:1997ad}:
\begin{equation}
    \begin{bmatrix}
        \rho_1(ku)\\
        \rho_2(ku)\\
        \rho_3(ku)
    \end{bmatrix} \equiv \frac{1}{2(ku)^2}\begin{bmatrix}
        2(ku)^2 & -4ku & 2\\
        -4(ku)^2 & 16ku & -20\\
        (ku)^2 & -10ku & 35
    \end{bmatrix}\begin{bmatrix}
        j_0(ku)\\
        j_1(ku)\\
        j_2(ku)
    \end{bmatrix}.
\end{equation}

In the \emph{short baseline limit}, where $u \ll 1/k$, the properties of the polarization basis tensors can be utilized:
\begin{equation}
    \int \frac{\d^2 \hat{k}}{{8\pi}}  \sum_A\hat\epsilon^{A}_{ab}(\hat{k})\hat\epsilon^{A}_{cd}(\hat{k})=\frac{1}{5}(\delta_{ac}\delta_{bd}+\delta_{bc}\delta_{ad})-\frac{2}{15}\delta_{ab}\delta_{cd},
\end{equation}
to simplify the ORF in Eq.~(\ref{eq:ORFS}) as
\begin{align}
    \gamma_{IJ} \xrightarrow{|\vec{x}_J-\vec{x}_I|f\ll1} 
    \frac{2}{5}R_I^{ab} R_J^{*ab}=\frac{2}{5}\mathrm{Tr}\lp R_I R_J^{\dagger}\rp=\sum_{m_I,m_J}\frac{2}{5} B_I^{m_I} B^{m_J*}_{J} \, \mathrm{Tr}\lp \hat{R}_{m_I} \hat{R}^{\dagger}_{m_J} \rp,
\end{align}
under the assumption that the response matrix is symmetric and traceless.

\begin{figure*}[t]
    \centering
    \includegraphics[width=0.27\textwidth]{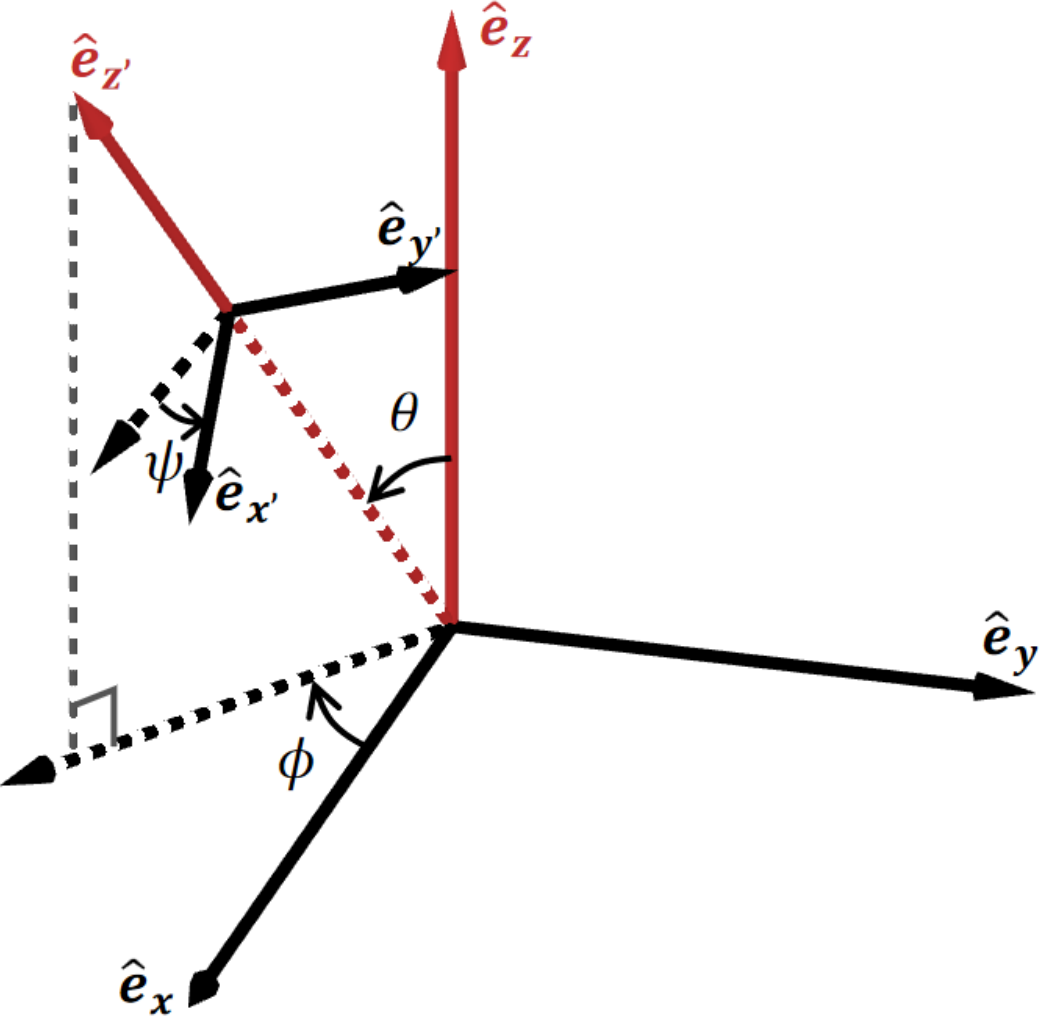}
    \includegraphics[width=0.63\textwidth]{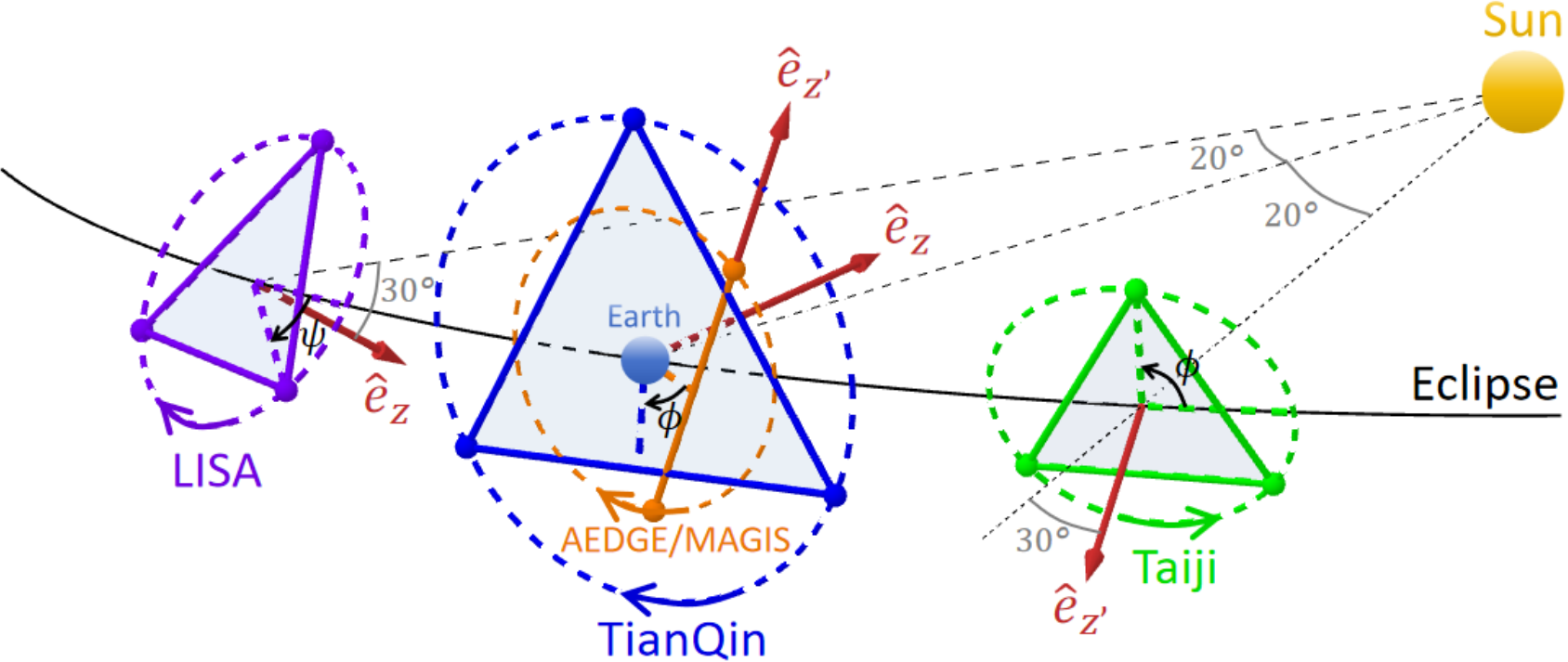}
    \caption{{\bf Left:}  Illustration of two frames with Cartesian coordinates $(\hat{e}_x, \hat{e}_y, \hat{e}_z)$ and $(\hat{e}_{x'}, \hat{e}_{y'}, \hat{e}_{z'})$, related through a rotation matrix defined by Euler angles $\Omega \equiv (\theta, \phi, \psi)$. $\theta$ represents the separation angle between $\hat{e}_z$ and $\hat{e}_{z'}$, while $\phi$ and $\psi$ are the rotation angles about these respective axes.
    {\bf Right:} Orbital and orientational configurations of the space missions considered: LISA-Taiji in heliocentric orbit and TianQin-AEDGE/MAGIS in Earth orbit. The normal vectors to the constellations for laser interferometers LISA, Taiji, and TianQin, as well as the baseline for the atom interferometer AEDGE/MAGIS, are represented by red arrows, denoted as $\hat{e}_z$ and $\hat{e}_{z'}$. The LISA-Taiji pair rotates in opposite directions when viewed from the Sun, with both $\phi$ and $\psi$ increasing linearly due to their self-rotation. Consequently, the combination $(\phi + \psi)$ grows with time, while $(\phi - \psi)$ remains constant.
    The orbital planes of TianQin and AEDGE/MAGIS are assumed to be aligned, resulting in perpendicular $\hat{e}_z$ and $\hat{e}_{z'}$ with $\phi$ modulating as the missions rotate.}
    \label{fig:config}
\end{figure*}

Each response matrix for the channels can be expressed using the bases defined in Eq.~(\ref{eq:RLB}) or Eq.~(\ref{eq:RCB}) for each respective frame. Calculating the ORF requires accounting for the relative orientation between detectors, parameterized by a rotation matrix $\mathcal{P}_a{}^b$ expressed via three Euler angles $\Omega \equiv (\theta, \phi, \psi)$:
\begin{equation}
\mathcal{P}(\Omega)=\mathcal{P}_z(-\phi)\mathcal{P}_y(\theta)\mathcal{P}_z(\psi)
=\begin{pmatrix}
    \cos\phi&-\sin\phi&0\\\sin\phi&\cos\phi&0\\0&0&1
\end{pmatrix}
\begin{pmatrix}
    \cos\theta&0&\sin\theta\\0&1&0\\-\sin\theta&0&\cos\theta
\end{pmatrix}
\begin{pmatrix}
    \cos\psi&\sin\psi&0\\-\sin\psi&\cos\psi&0\\0&0&1
\end{pmatrix}.
\end{equation}
Under this transformation, the response matrix $R'_J$, initially defined in its own Cartesian frame using axes $(\hat{e}_{x'}, \hat{e}_{y'}, \hat{e}_{z'})$, transforms to $R_J = \mathcal{P} R'_J \mathcal{P}^\dagger$ in the $I$-th detector’s frame with axes $(\hat{e}_{x}, \hat{e}_{y}, \hat{e}_{z})$. 
As shown in the left panel of Fig.~\ref{fig:config}, $\theta$ is the separation angle between the detectors' $\hat{e}_z$-axes, and $\phi$ and $\psi$ are rotation angles about these respective axes.

Working in the circular polarization basis as defined in Eq.~(\ref{eq:RCB}), the angular correlation between two channels in their respective bases, $R_I= \Sigma_m B^m_I\hat{R}_m$ and $R'_J= \Sigma_m B^m_J\hat{R}_m$, with their relative orientation described by $\mathcal{P}(\Omega)$, can be analytically determined as follows~\cite{Chu:2020qiw}:
\begin{equation}
\gamma_{IJ}(\Omega)=\sum_{m,n} \frac{2}{5} B_I^{m} B^{n*}_J\,
\mathrm{Tr}\lp \hat{R}_{m} \mathcal{P} \hat{R}_{n}^{\dagger} \mathcal{P}^\dagger\rp = \sum_{m,n} \frac25 B_I^{m} B^{n*}_J\, D^{j=2}_{m,n}(-\psi,\theta,\phi).
\end{equation}
Here, $D^j_{m,n}(-\psi,\theta,\phi)$ represents the spin-$2$ Wigner D-matrix~\cite{wigner1931gruppentheorie}, which quantifies the overlap between the modes $m$ and $n$ of the $j=2$ representation, as defined in their respective frames, which differ by the rotation $\mathcal{P}(\Omega)$.

The magnitude of the Wigner D-matrix depends solely on $\theta$, while its phase is influenced by $\phi$ and $\psi$. More specifically, the azimuthal angles $\psi$ and $\phi$ contribute phase factors, expressed as $\mathcal{P}^\dagger_z(-\phi)\hat R_m\mathcal{P}_z(-\phi)=e^{im\phi}\hat R_m$ and $\mathcal{P}_z(\psi)\hat R^{\dagger}_n\mathcal{P}^\dagger_z(\psi)=e^{-in\psi}\hat R^{\dagger}_n$. Therefore, the Wigner D-functions can be formulated as:
\begin{align}
    D^{j}_{m,n}(\psi,\theta,\phi)= \me^{i(m\psi+n\phi)}D^{j}_{m,n}(0,\theta,0).
\end{align}
The polar angle $\theta$ determines the amplitude as: $D^{j=2}_{m,n}(0,\theta,0)=\tr\ls\hat{R}_{m}\mathcal{P}_y(\theta)\hat{R}^{\dagger}_n\mathcal{P}^\dagger_y(\theta)\rs$,
resulting in six independent components with respect to the azimuthal numbers:
\begin{subequations}
\begin{align}
    D^{j=2}_{0,0}(0,\theta,0)&=\frac{1}{4}(1+3\cos{2\theta}), \\
    D^{j=2}_{1,0}(0,\theta,0)&=\sqrt{\frac{3}{2}}\cos\theta\sin{\theta}, \\
    D^{j=2}_{2,0}(0,\theta,0)&=\frac{1}{2} \sqrt{\frac{3}{2}} \sin ^2\theta, \\
    D^{j=2}_{1,1}(0,\theta,0)&=\frac{1}{2}(\cos{\theta}+\cos{2\theta}), \\
    D^{j=2}_{2,1}(0,\theta,0)&=\frac{1}{2}(1+\cos\theta)\sin\theta, \\
    D^{j=2}_{2,2}(0,\theta,0)&=\cos^4{\frac{\theta}{2}}.
\end{align}
\end{subequations}
Other components are related by the symmetries:
\begin{align}
   D^{j=2}_{m,n}(0,\theta,0)&=(-1)^{m+n}D^{j=2}_{n,m}(0,\theta,0)=(-1)^{m+n}D^{j=2}_{-m,-n}(0,\theta,0) \\
    &=(-1)^{n}D^{j=2}_{m,-n}(0,\theta-\pi,0)=(-1)^{m}D^{j=2}_{-m,n}(0,\theta-\pi,0).
\end{align}

For instance, correlations between two atom interferometers scale as $D^2_{00} = (1 + 3 \cos 2 \theta)/4$, similar to the Hellings-Downs curve, excluding the redshift factor. 

Since the missions considered in this study exhibit simple responses corresponding to one of the bases in Eq.~(\ref{eq:RLB}), the relevant coefficients $B_m^I$ are taken as unity for simplicity in the following.

\section{Statistics}\label{sec:statistics}

This section outlines the analytical forecasts for detecting the SGWB and identifying its quadrupolar nature. Particularly, we analyze the timescales required for both discovery and quadrupolar identification.

Consider a network of GW detectors operating simultaneously, comprising $N_{\rm ch}$ channels with overlapping frequency coverage. The total observation time $T_{\rm tot}$ is divided into multiple segments of equal duration $T_{\rm seg}$, which is much shorter than the timescale over which detector configurations evolve. Within each segment, synchronized observations between pairs of detectors allow the correlations between any two response channels to be treated as constant and computed using Eq.~(\ref{avg}). The quadrupolar information is encoded in the ORF matrix $\gamma_{IJ}$, which varies from segment to segment as the detector orientation $\Omega$ changes.

Within each segment, data are transformed into the frequency domain and labeled as ${d_I^{k,\alpha}}$, where the integers $I$, $k$, and $\alpha$ represent the channel, the frequency bin with central frequency $k/T_{\rm seg}$, and the segment index, respectively. Assuming the noise in different channels is independent and Gaussian, the likelihood for model parameters $\mathcal{O}$ given the dataset $\boldsymbol{d}$ can be written as:
\begin{equation}\label{eq:SMlikelihood}
    \mathcal{L}(\boldsymbol{d} \,\vert \mathcal{O} )=\prod_{k,\alpha} \frac{1}{\mathrm{Det}\left[\pi C^{k,\alpha}(\mathcal{O})\right]}\exp \left\{-\sum_{I,J} d_I^{k,\alpha} \left[C^{k,\alpha}(\mathcal{O})\right]^{-1}_{IJ} (d_J^{k,\alpha})^{\dagger} \right\},
\end{equation}
where the covariance matrix is defined as:
\begin{equation}
    \left[C^{k,\alpha}(\mathcal{O})\right]_{IJ} \equiv \frac{1}{2}\left[\delta_{IJ} S_{n_I}^k+ \gamma_{IJ}(\Omega_{IJ}^{\alpha}) S_h^k \right]. 
    \label{eq:CM}
\end{equation}
Here, $S_{n_I}^k$ represents the noise PSD of channel $I$, and $S_h^k$ is the PSD of the SGWB. $\Omega_{IJ}^{\alpha}$ indicates the Euler angles describing the relative configuration between channels $I$ and $J$ for segment $\alpha$, and $\gamma_{IJ}(\Omega_{IJ}^{\alpha})$ is the corresponding ORF, as defined in Eq.~(\ref{eq:ORFS}). The subsequent discussion will focus primarily on the ratio between the two terms in Eq.~(\ref{eq:CM}), with the auto-correlation normalized to $\gamma_{II}=2/5$ and the noise PSD adjusted accordingly.

The capability of the detector network to estimate model parameters is quantified by the information matrix~\cite{doi:10.1098/rsta.1922.0009}:
\begin{equation}
    \mathcal{I}_{ij} \equiv - \left\langle \frac{ \partial^2 \mathrm{ln} \, \mathcal{L}}{\partial \mathcal{O}_i \partial \mathcal{O}_j}\right\rangle \bigg|_{\mathcal{O}=\mathcal{O}_{\mathrm{truth}}},
\end{equation}
where $\mathcal{O}_{\mathrm{truth}}$ represents the true parameters. For any given model parameter $\mathcal{O}_i$, the variance of its estimator is expressed as:
\begin{equation}
    \sigma^2_i \equiv (\mathcal{I})^{-1}_{ii},
\end{equation}
and the corresponding signal-to-noise ratio (SNR) is defined as:
\begin{equation}
    \mathrm{SNR}^2_i \equiv \mathcal{O}^2_{i,\,\mathrm{truth}}/\sigma^2_i.
    \label{eq: SNR_orig}
\end{equation}

To detect SGWB-induced power excess, the relevant measure is the signal-to-noise ratio (SNR) for $\ln S_h^k$, treated as the model parameter, evaluated for each frequency bin $k$ across a detector pair. We define dimensionless ORFs $\tilde{\gamma}_{IJ}\equiv (1-\delta_{IJ})\gamma_{IJ}/\sqrt{\gamma_{II} \gamma_{JJ}}$ to characterize the relative strength of cross-correlations compared to auto-correlations. Additionally, we introduce $\xi^k_I \equiv \gamma_{II} S_h^k / S_{n_I}^k$ to represent the the PSD ratio between signal and noise in the auto-correlation of each channel. Two signal regimes are considered: the {\emph{weak signal limit}} ($\xi_I^k \ll 1$), where measurement uncertainty is dominated by instrumental noise, and the {\emph{strong signal limit}} ($\xi_I^k \gg 1$), where the intrinsic stochastic fluctuations of the SGWB dominate.

In these two limits, the SNR at a given frequency bin $k$, denoted by $\mathrm{SNR}_k^2$, simplifies to:
\begin{equation}
\begin{aligned}
    \mathrm{SNR}^2_k
    &\approx
\begin{cases}
    \frac{T_{\rm tot}}{T_{\rm seg}} \left( \sum_I (\xi_I^k)^2 + \sum_{I,J} \xi_I^k \xi_J^k \langle|\tilde{\gamma}_{IJ}|^2 \rangle_{\Omega} \right), \qquad \xi_I^k \ll 1, \\
     \frac{N_{\rm ch}\,T_{\rm tot}}{T_{\rm seg}}, \qquad \xi_I^k \gg 1,
\end{cases}
\end{aligned}
\label{eq:SNR}
\end{equation}
where $\langle \cdots \rangle_\Omega$ denotes the average over channel configurations due to changing Euler angles, and $T_{\rm tot}/T_{\rm seg}$ is the number of independent data segments. In the {\emph{weak signal limit}}, the first and second terms represent contributions from auto-correlations and cross-correlations, respectively, with quadrupolar information encoded solely in the cross-correlation terms. In the {\emph{strong signal limit}}, each channel contributes equally. Notably, cross-terms between different channels of a single detector vanish due to the orthogonality of the response basis.

After detecting potential signals from the SGWB, it is crucial for a network of GW detectors to confirm their quadrupolar nature in order to distinguish them from instrumental or environmental noise. A quantitative criterion for assessing this feature can be formulated using the log-likelihood ratio ($\lambda_{\rm LR}$) between the full correlation model and a model including only auto-correlations:
\begin{equation}
    \lambda_{\rm LR} \equiv 2\left[\ln \mathcal{L}(\boldsymbol{d} \,\vert \mathcal{O}_{\rm truth} )-\ln \mathcal{L}_{\rm auto}(\boldsymbol{d} \,\vert \mathcal{O}_{\rm truth} ) \right],
    \label{eq: LR_def}
\end{equation}
where $\mathcal{L}_{\rm auto}$ is the likelihood computed using only auto-correlations, analogous to Eq.~(\ref{eq:SMlikelihood}) but with a covariance matrix modified to
\begin{equation}
    \left[C^{k,\alpha}_{\rm auto}(\mathcal{O})\right]_{IJ} \equiv \frac{1}{2} \delta_{IJ} \lp S_{n_I}^k+  \gamma_{II}S_h^k \rp.
\end{equation}

In each frequency bin, the ensemble-averaged log-likelihood ratio simplifies in the two signal limits to
\begin{equation}
\begin{aligned}
    \langle \lambda_{\rm LR}^k \rangle &= 2 \sum_{\alpha}\left\{ \ln \mathrm{Det} \left[ C_{\rm auto}^{k,\alpha}(\mathcal{O}_{\rm truth})\right] - \ln \mathrm{Det} \left[ C^{k,\alpha}(\mathcal{O}_{\rm truth})\right] \right\} \\
    &\approx
    \begin{cases}
    \frac{T_{\rm tot}}{T_{\rm seg}} \sum_{I,J} \xi_I^k \xi_J^k \langle|\tilde{\gamma}_{IJ}|^2 \rangle_{\Omega} , \qquad \xi_I^k \ll 1, \\
    -\frac{2\,T_{\rm tot}}{T_{\rm seg}}  \langle\mathrm{ln} \, \mathrm{Det}\left[\mathbf{I}_{N_{\rm ch}}+\tilde{\gamma}\right] \rangle_{\Omega}, \qquad \xi_I^k \gg 1,
\end{cases}
\end{aligned}
\label{eq:LR}
\end{equation}
where $\tilde{\gamma}$ denotes the $N_{\rm ch} \times N_{\rm ch}$ matrix of normalized cross-correlation terms $\tilde{\gamma}_{IJ}$, and $\mathbf{I}_{N_{\rm ch}}$ is the identity matrix. In deriving this result, the ensemble average simplifies the exponential term in Eq.~(\ref{eq:SMlikelihood}) to $\exp(-N_{\rm ch})$, leaving only the generalized variances of the full and auto-correlations. Comparing with the SNR expression in Eq.~(\ref{eq:SNR}), the log-likelihood ratio in the \emph{weak signal limit} is exactly given by the cross-correlation contribution, and thus directly serves as a statistical criterion for confirming the quadrupolar nature of the SGWB. In the \emph{strong signal limit}, it depends solely on the full correlation matrix and becomes independent of instrumental noise, reflecting intrinsic SGWB fluctuations.

By comparing Eq.~(\ref{eq:SNR}) and Eq.~(\ref{eq:LR}), it becomes evident that the duration required to confirm the quadrupolar nature of the SGWB, $T_Q$, generally exceeds the time required for the initial detection of a power excess, $T_D$. We now compare these two timescales. Consider a power-law SGWB spectrum defined as $S_h(f) = A_{\rm GW}^2 (f/f_0)^{2\Gamma} / f$, where $A_{\rm GW}$ denotes the strain amplitude, $f_0$ is a reference frequency, and $\Gamma$ characterizes the spectral slope. We treat $\ln A_{\rm GW}$ as the model parameter when computing the SNR via Eq.~(\ref{eq: SNR_orig}), assuming $\Gamma$ is fixed. The total SNR is obtained by summing over frequency bins: $\mathrm{SNR}^2 = \Sigma_k \mathrm{SNR}^2_k$. The required observation time for detection at $90\%$ confidence is defined by $\mathrm{SNR}^2(T_D) = 2.71$.

Likewise, the log-likelihood ratio for identifying the quadrupolar nature, as given in Eq.~(\ref{eq: LR_def}), is summed across frequencies as $\langle \lambda_{\rm LR} \rangle \equiv \Sigma_k \langle \lambda_{\rm LR}^k \rangle$. The threshold condition for quadrupolar confirmation is set to match the discovery threshold: $\langle \lambda_{\rm LR}(T_Q) \rangle = \mathrm{SNR}^2(T_D) = 2.71$. Substituting the expressions from Eqs.~(\ref{eq:SNR}) and (\ref{eq:LR}) into these definitions yields the ratio:
\begin{equation}
    \frac{T_Q}{T_D}=
\begin{cases}
     1+\frac{\sum_{k^{\prime},L}  (\xi_L^{k^{\prime}})^2}{\sum_{k,I,J} \xi_I^k \xi_J^k \langle|\tilde{\gamma}_{IJ}|^2 \rangle_{\Omega}}, \qquad \xi_I^k \ll 1, \\ \noalign{\vskip5pt}
     -\frac{N_{\rm ch}}{2\,\langle\mathrm{ln} \, \mathrm{Det}\left[\mathbf{I}_{N_{\rm ch}}+\tilde{\gamma}\right] \rangle_{\Omega}}, \qquad \xi_I^k \gg 1,
\end{cases}
\label{eq:TR}
\end{equation}
assuming all relevant frequency bins fall within either the \emph{weak signal limit} or the \emph{strong signal limit}. The ratio $T_Q / T_D$ depends sensitively on detector sensitivity and the orbital configuration of the detector network across frequency.

Note that the above formulas apply to both \emph{short} and \emph{long baseline} regimes; in the latter case, the full ORF expression in Eq.~(\ref{full}) should be used.


\section{Identifying Quadrupolar Nature through Orbital Motion of Space Missions}\label{sec:IQNS}

The quadrupolar signature of the SGWB can be revealed by scanning the ORFs across the configuration space defined by Euler angles. While PTAs and astrometric observations benefit from large numbers of baselines—approximately $10^2$ and $10^9$, respectively—current and near-future terrestrial and space-based GW missions operate with only a limited number of detector pairs within a given frequency band. Nevertheless, the orbital motion and self-rotation of space-based detectors enable effective exploration of the ORF configuration space, inducing time-dependent correlation modulations that reflect the SGWB’s quadrupolar structure.

In this section, we examine representative examples of laser and atom interferometers aboard space missions, whose orbital and orientational dynamics naturally sample the relevant configuration space. In particular, we focus on two detector pairs: LISA-Taiji, operating in heliocentric orbit, and TianQin-AEDGE/MAGIS, in geocentric orbit. Their orbital and orientation configurations are illustrated in the right panel of Fig.~\ref{fig:config}. These examples serve to evaluate the capability of such systems to detect the SGWB and identify its quadrupolar nature.

In practice, the timescale over which the Euler angles vary—approximately one year for heliocentric orbits and one month for Earth orbits—greatly exceeds the duration required to resolve the lowest frequency bins (on the order of $10^4$~s). This allows synchronized observations between detector pairs to be divided into multiple time segments. As these missions exhibit simple responses corresponding to individual basis elements in Eqs.~(\ref{eq:RCB}), we set the relevant coefficients $B_m^I$ to unity for simplicity.

Correlations involving terrestrial detectors operating in overlapping frequency bands can be similarly analyzed by incorporating their respective orbital and orientation parameters.

The benchmark SGWB we consider originates from stellar-mass compact binary populations, including BHs and neutron stars, characterized by a spectrum $S_h(f)\equiv h_c(f)^2/f$ with $h_c = 8.1\times10^{-25}\times(f/\mathrm{Hz})^{-2/3}$ representing the characteristic strain, across frequencies from $10^{-4}$~Hz to $10^2$~Hz~\cite{KAGRA:2021duu}. This irreducible background is in the weak signal limit for the four missions considered.

\subsection{Space Missions}

\subsubsection{Laser Interferometers}

Laser interferometers, such as LIGO and LISA, precisely measure GWs through the interference patterns of light beams. LIGO functions as an equal-arm Michelson interferometer, which splits a laser beam into two perpendicular paths. Each beam travels to a mirror, reflects back, and is then recombined to interfere with each other. Slight changes in the distance to one of the mirrors, caused by GWs, affect the phase relationship between the two beams, thereby altering the interference pattern. This effect is proportional to the contraction of both strain indices with the same distance vector. Assuming the arm directions as $\hat{e}_x$ and $\hat{e}_y$, the response matrix for LIGO is defined as:
\be R_{\rm LIGO}=\hat{R}_{T+}=\frac{1}{\sqrt{2}}(\hat{R}_{+2}+\hat{R}_{-2}). \label{eq:LIGO}\ee

\begin{figure}[htbp]
    \centering
    \includegraphics[width=0.4\textwidth]{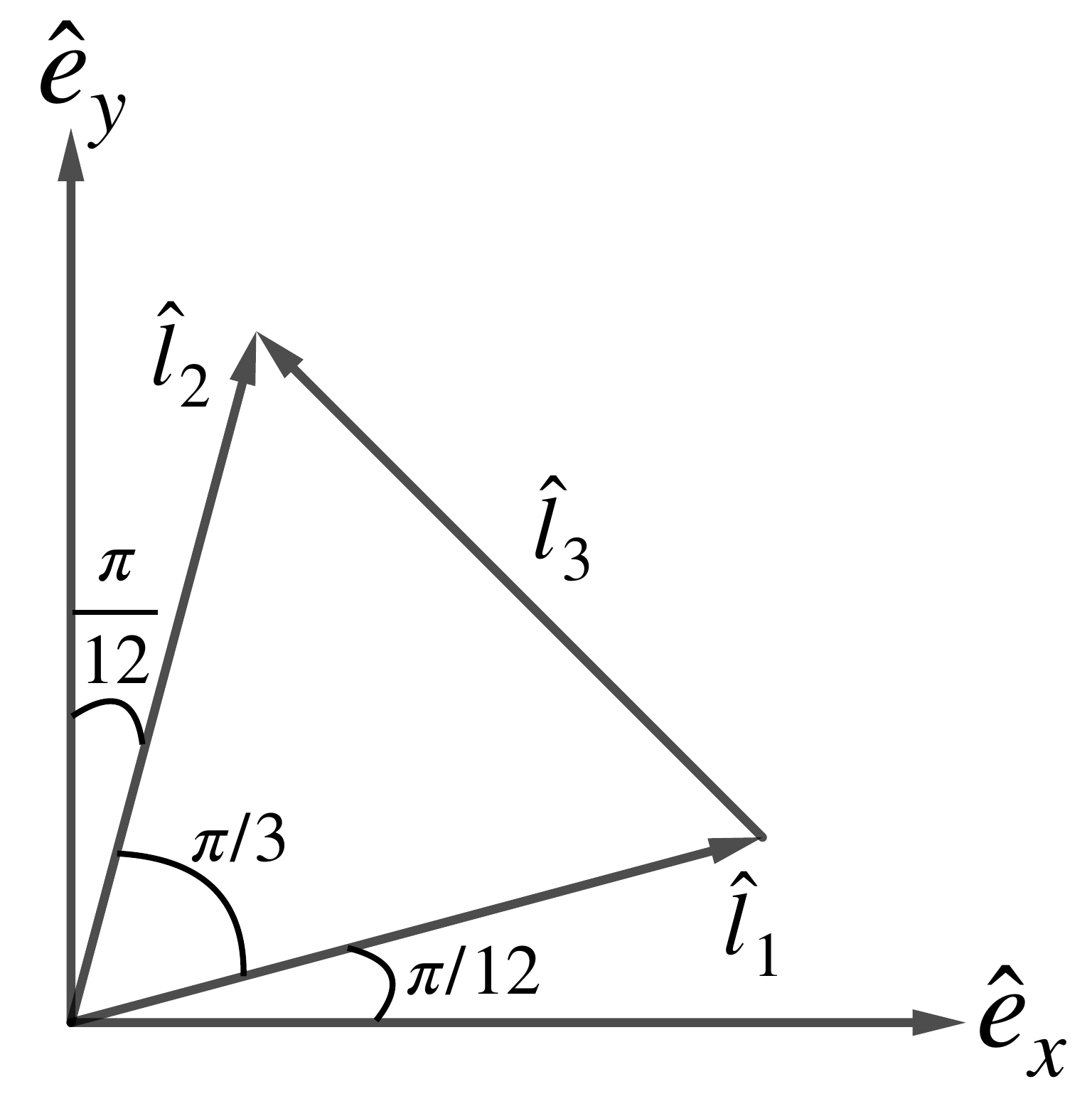}
    \caption{Configuration of a LISA-like GW detector constellation, depicting three spacecraft positioned at each vertex of an almost equilateral triangle. The unit directional vectors $\hat{l}_i$ represent the paths of the laser beams connecting the satellites. The $x-y$ coordinate system is chosen such that $\hat l_i=\cos\gamma_i\hat e_{x}+\sin\gamma_i\hat e_{y}$, where $i=1,2,3$ and $\gamma_i=\pi/12+(i-1)\pi/3$. The constellation features two independent response channels with response matrices $\hat{R}^{T+}$ and $\hat{R}^{T\times}$.
    }
    \label{fig:LISAconfig}
\end{figure}

A LISA-like space-based detector comprises three satellites positioned at each vertex of an almost equilateral triangle. Each satellite houses two free-falling test masses, sending and receiving light to and from the nearby satellites, thus forming an interference pattern. By comparing the interference patterns of the two light paths, one can derive three combinations of response matrices: $X=\hat l_1\hat l_1-\hat l_2\hat l_2, Y=\hat l_2\hat l_2-\hat l_3\hat l_3,$ and $Z=\hat l_3\hat l_3-\hat l_1\hat l_1$, where $\hat{l}_i$ represents the unit directional vectors of the three beams, as illustrated in Fig.~\ref{fig:LISAconfig}. In the center of mass frame of the triangle, defining $\hat{e}_x$ and $\hat{e}_y$ allows $\hat l_i$ to be expressed as $\hat l_i=\cos\gamma_i\hat e_{x}+\sin\gamma_i\hat e_{y}$, where $i=1,2,3$ and $\gamma_i=\pi/12+(i-1)\pi/3$. This definition aligns $X$ with LIGO’s response matrix in Eq.~(\ref{eq:LIGO}) scaled by a factor of ${\sqrt{3}}/{4}$. Since only two of the three channels are independent, we can reconfigure the Michelson combinations $X$, $Y$, and $Z$ into three optimal combinations $A$, $E$, and $T$~\cite{Prince:2002hp,Gair:2012nm}:
\begin{align}
	A&=\frac{2}{3}(Y-Z)=\hat{R}^{T+}=\frac{1}{\sqrt{2}}(\hat{R}_{+2}+\hat{R}_{-2}),\\
	E&=\frac{2}{3\sqrt{3}}(Y+Z-2X)=\hat{R}_{T\times}=\frac{1}{i\sqrt{2}}(\hat{R}_{+2}-\hat{R}^{-2}),\\
	T&=\frac{1}{3}(X+Y+Z)=0.
 \label{LISAportal}
\end{align}
Here, the $A$ and $E$ channels can be treated as independent, while $T$ is a null channel insensitive to GW signals, useful for calibrating measurement noise.

We examine three space-based laser interferometer missions: LISA and Taiji, both operating in heliocentric orbits, and TianQin, which orbits the Earth. The main sources of noise in space-based laser interferometers stem from two categories: optical metrology system (OMS) noise, which includes shot noise, and acceleration noise related to the test masses~\cite{Babak:2021mhe,Colpi:2024xhw}. For LISA and Taiji, the corresponding PSDs,  $S_{\mathrm{OMS}}(f)$  and  $S_{\mathrm{acc}}(f)$, are expressed as~\cite{Babak:2021mhe,Ren:2023yec}:
\begin{equation} \begin{split}
    S_{\mathrm{OMS}}(f)= \ &(2\pi f)^2S_x\ls1+\lp\frac{2\,\rm{mHz}}{f}\rp^4\rs,\\
    S_{\mathrm{acc}}(f)= \ &\frac{S_a}{(2\pi f)^2}\ls1+\lp\frac{0.4\,\rm{mHz}}{f}\rp^2\rs\ls1+\lp\frac{f}{8\,\rm{mHz}}\rp^4\rs.
\end{split}\end{equation}
The relevant parameters are:
\begin{align}
    \text{LISA: }& S_x^{1/2}=15\times 10^{-12}\,\mathrm{m}\cdot\mathrm{Hz}^{-1/2}, S_a^{1/2}=3\times 10^{-15}\,\mathrm{m}\cdot\mathrm{s}^{-2}\cdot\mathrm{Hz}^{-1/2}, L=2.5\times 10^6 \,\mathrm{km},
    \\
    \text{Taiji: }& S_x^{1/2}=8\times 10^{-12}\,\mathrm{m}\cdot\mathrm{Hz}^{-1/2},~\,  S_a^{1/2}=3\times 10^{-15}\,\mathrm{m}\cdot\mathrm{s}^{-2}\cdot\mathrm{Hz}^{-1/2}, L=3\times 10^6 \,\mathrm{km},
\end{align}
where $L$ denotes the arm length. The total noise PSD for one channel, whether $A/\hat{R}^{T+}$ or $E/\hat{R}^{T\times}$, is given by~\cite{Babak:2021mhe}:
\begin{equation}
    S_n(f)=\frac{8}{3}\frac{S_{\mathrm{OMS}}(f)+(3+\cos4\pi fL)S_{\mathrm{acc}}(f)}{(2\pi fL)^2}\ls1+0.6(2\pi f L)^2\rs,
    \label{TLformula}
\end{equation}
where the factor $8/3$ is introduced to normalize the auto-correlation to $\gamma_{II} = 2/5$ as compared with the literature~\cite{Babak:2021mhe}, and the term $\ls1+0.6(2\pi f L)^2\rs$ accounts for the deviations from the \emph{small-antenna limit}, as discussed in Refs.~\cite{Babak:2021mhe,Lu_2019}. We consider a frequency range for the two between $10^{-4}$\,Hz and $10^{-1}$\,Hz, where acceleration noise dominates at frequencies below $10^{-3}$\,Hz and OMS noise at higher frequencies.

For TianQin, the noise PSD can be approximated as~\cite{TianQin:2015yph}:
\begin{equation}
    S_n(f)=\frac{8}{3}\frac{1}{L^2}\ls S_x+\frac{(3+\cos4\pi fL)S_a}{(2\pi f)^4}\lp1+\frac{10^{-4}\,\mathrm{Hz}}{f}\rp\rs\ls1+0.6(2\pi f L)^2\rs,
    \label{TQformula}
\end{equation}
with 
\be S_x^{1/2}=1\times 10^{-12}\,\mathrm{m}\cdot\mathrm{Hz}^{-1/2}, S_a^{1/2}=1\times 10^{-15}\,\mathrm{m}\cdot\mathrm{s}^{-2}\cdot\mathrm{Hz}^{-1/2}, L=1.7\times 10^5 \,\mathrm{km}.\ee
Given the shorter arm length compared to LISA and Taiji, TianQin’s sensitive frequency range is slightly higher, spanning $10^{-3}$\,Hz to $1$\,Hz.

\begin{figure}[htbp]
    \centering
    \includegraphics[width=0.75\textwidth]{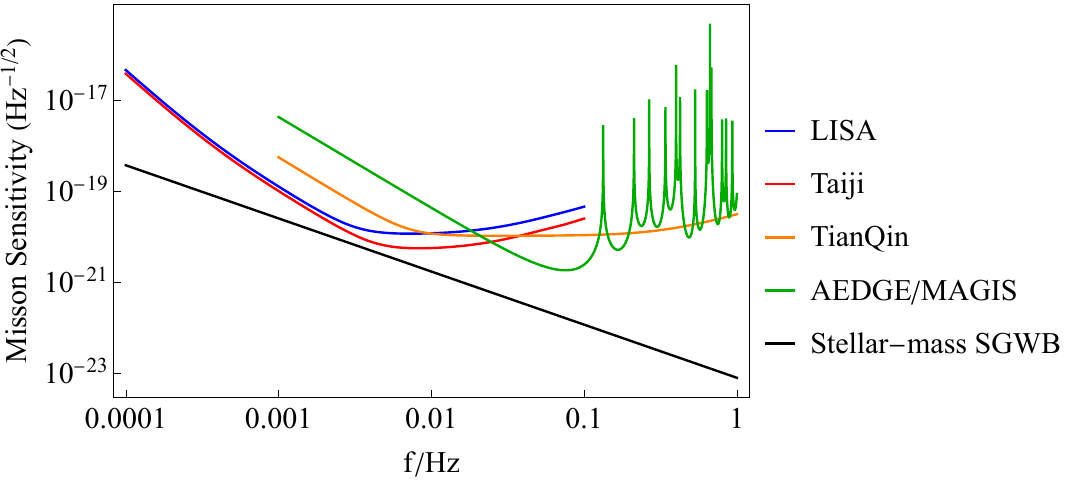}
    \caption{Mission sensitivity curves, $\sqrt{S_n/(N_{\rm ch}\gamma_{II})}$, for the four missions-LISA, Taiji, TianQin and AEDGE/MAGIS-are presented. The number of channels is $N_{\rm ch} = 2$ for laser interferometers and $N_{\rm ch} = 1$ for the atom interferometer, with the auto-correlation factor $\gamma_{II} = 2/5$. The LISA-Taiji pair covers a sensitivity frequency range from $10^{-4}$\,Hz to $10^{-1}$\,Hz, whereas TianQin-AEDGE/MAGIS spans $10^{-3}$\,Hz to $1$\,Hz. The SGWB spectrum $\sqrt{S_h}$ from stellar-mass compact binary populations, characterized by $S_h(f) \equiv h_c(f)^2/f$, where $h_c = 8.1 \times 10^{-25} \times (f/\mathrm{Hz})^{-2/3}$~\cite{KAGRA:2021duu}, is shown for comparison. This indicates that all four missions operate within the \emph{weak signal limit} across their frequency ranges.}
    \label{figSM:sensitivities}
\end{figure}

In Fig.~\ref{figSM:sensitivities}, we present the mission sensitivities, represented as $\sqrt{S_n/(N_{\rm ch}\gamma_{II})}$, where $N_{\rm ch} = 2$ for laser interferometers. For comparison, we also illustrate the benchmark SGWB spectrum, $\sqrt{S_h}$, which originates from compact binary populations~\cite{KAGRA:2021duu}.

\subsubsection{Long-baseline Atom Interferometers}
Atom interferometers harness a sequence of laser pulses to manipulate coherent ensembles of atoms, effectively splitting and then recombining their wave packets to generate interference patterns~\cite{Dimopoulos:2007cj,Dimopoulos:2008sv}. During operation, these atom wave packets traverse two distinct paths: one remains in the ground state, while the other is excited by the laser pulses. A phase difference accumulates between the two paths, influenced by both the arrival time of the laser pulses and the free-fall evolution of the atoms. This phase difference becomes sensitive to GWs passing through the setup. The GW-induced phase shift is directly proportional to the distances traveled by both the atoms and the laser pulses. Practically, expanding the space available for ultra-cold atom experiments is challenging. A viable approach is to implement a long laser baseline setup, where common laser pulses trigger two separate atom interferometers situated a significant distance apart. In such configurations, exemplified by space missions like AEDGE~\cite{AEDGE:2019nxb} and MAGIS~\cite{Graham:2017pmn}, the GW-induced phase shift is predominantly influenced by the laser’s travel path. Consequently, the GW response is largely determined by the projection of the strain tensor onto the baseline vector, resulting in a traceless response matrix:
\begin{equation}
    R_{\rm AI}=\hat{R}_{SL}=\hat{R}_0=\sqrt{\frac{3}{2}}\lp\hat e_z\otimes \hat e_z - \frac{\mathbf{I}}{3}\rp,
\end{equation}
where $\hat{e}_z$ represents the baseline direction.

In long-baseline atom interferometers such as AEDGE and MAGIS, a single optical path exists, which avoids the acceleration noise typically associated with back-reaction from test masses. The predominant source of noise in these configurations is shot noise, inherent to the atom phase readout, denoted as $S_{\mathrm{atom}}$. This translates directly into the measurement noise for the system within the \emph{small-antenna limit}~\cite{Graham:2016plp,Graham:2017pmn,Wang:2021hrg}:
\begin{equation}
    S_n(f)=\frac{3}{2}\frac{S_{\mathrm{atom}}}{\frac{2f_A}{f}\frac{\sin(2\pi f Q T_{\mathrm{int}})}{\cos(\pi f T_{\mathrm{int}})}\sin\ls\pi f T_{\mathrm{int}}-\pi f(N-1)L\rs\sin\left(\pi f NL\right)}.
    \label{eqSM:MAformula2}
\end{equation}
Here, $f_A$ is the atom transition frequency, $Q$ is the number of atom recombinations until the interferometry phase measurement, $N$ is the number of laser pulses used for exciting/de-exciting the atoms during one recombination period, and $T_{\mathrm{int}}$ is the duration of one recombination period. The additional factor $3/2$ compared with the literature~\cite{Graham:2016plp,Graham:2017pmn,Wang:2021hrg} is for normalization of the auto-correlation. For AEDGE/MAGIS, the parameters using $^{87}\mathrm{Sr}$ are as follows~\cite{Graham:2016plp,Graham:2017pmn}:
\be S_{\mathrm{atom}}^{1/2}=10^{-5}\,\mathrm{Hz}^{-1/2},\ f_A=6.83\times10^{13}\,\mathrm{Hz},\ Q=1,\ N=20,\ T_{\rm int} = 7.5\,\textrm{s},\ L=4.4\times10^4\,\mathrm{km}.\ee
The sensitive curve $\sqrt{S_n/(N_{\rm ch}\gamma_{II})}$, where $N_{\rm ch} = 1$, is also shown in Fig.~\ref{figSM:sensitivities} to compare with laser interferometers and the benchmark SGWB, featuring a frequency range similar to TianQin.

\subsection{LISA-Taiji Synergy}

Each LISA-like detector features two independent channels, $A=\hat{R}_{T+}$ and $E=\hat{R}_{T\times}$. Consequently, there are four possible combinations of correlations between two detectors. In the \emph{short baseline limit}, where the GW wavelength greatly exceeds the separation between LISA and Taiji (frequency range $f\ll 2.9\times10^{-3}$~Hz~\cite{Joffre:2021hne}), the ORFs parameterized by three Euler angles are given below
\begin{subequations}
\begin{align}
\tilde{\gamma}_{\rm{LT}}^{++}&=2(D^2_{2,2}+D^2_{2,-2}+D^2_{-2,2}+D^2_{-2,-2})=\cos2(\phi-\psi)\cos^4\frac{\theta}{2}+\cos2(\phi+\psi)\sin^4\frac{\theta}{2},\label{gamma221}\\
\tilde{\gamma}_{\rm{LT}}^{\times\times}&=2(D^2_{2,2}-D^2_{2,-2}-D^2_{-2,2}+D^2_{-2,-2})=\cos2(\phi-\psi)\cos^4\frac{\theta}{2}-\cos2(\phi+\psi)\sin^4\frac{\theta}{2},\\
\tilde{\gamma}_{\rm{LT}}^{+\times}&=2i(D^2_{2,2}-D^2_{2,-2}+D^2_{-2,2}-D^2_{-2,-2})=\sin2(\phi-\psi)\cos^4\frac{\theta}{2}-\sin2(\phi+\psi)\sin^4\frac{\theta}{2},\\
\tilde{\gamma}_{\rm{LT}}^{\times +}&=-2i(D^2_{2,2}+D^2_{2,-2}-D^2_{-2,2}-D^2_{-2,-2})=-\sin2(\phi-\psi)\cos^4\frac{\theta}{2}-\sin2(\phi+\psi)\sin^4\frac{\theta}{2}.\label{gamma222}
\end{align}
\end{subequations}
Here, `L' and `T' refer to LISA and Taiji, respectively. The correlations involving switches from $+$ to $\times$ differ simply by shifting either $\psi\rightarrow\psi+\pi/4$ or $\phi\rightarrow\phi+\pi/4$. The cross-correlation matrix for this pair is defined as follows:
\begin{equation}
    \tilde{\gamma}=
    \begin{pmatrix}
        0 & 0 & \tilde{\gamma}_{\mathrm{LT}}^{++} & \tilde{\gamma}_{\mathrm{LT}}^{+\times} \\
        0 & 0 & \tilde{\gamma}_{\mathrm{LT}}^{\times+} & \tilde{\gamma}_{\mathrm{LT}}^{\times\times} \\
        \tilde{\gamma}_{\mathrm{LT}}^{++} & \tilde{\gamma}_{\mathrm{LT}}^{+\times} & 0 & 0 \\
        \tilde{\gamma}_{\mathrm{LT}}^{\times+} & \tilde{\gamma}_{\mathrm{LT}}^{\times\times} & 0 & 0 
    \end{pmatrix}.
\end{equation}

The orbital configurations of these missions are depicted in the right panel of Fig.~\ref{fig:config}. LISA trails Earth by $20^\circ$, and Taiji leads by the same margin, with arm lengths of $2.5\times10^6\,\mathrm{km}$ and $3\times 10^6\,\mathrm{km}$, respectively. Both constellations are inclined at $60^\circ$ relative to the ecliptic plane and complete an annual rotation as they orbit the sun. 

In heliocentric orbit, the normal vectors to the orbital planes of LISA and Taiji, corresponding to the $\hat{e}_z$ direction in their respective Cartesian frames, can be parameterized as:
\begin{equation}
    \hat n_{\rm{L}}=\lp\sin\Theta_{\rm{L}}\cos\Phi_{\rm{L}}, \sin\Theta_{\rm{L}}\sin\Phi_{\rm{L}}, \cos\Theta_{\rm{L}}\rp, \quad
    \hat n_{\rm{T}}=\lp\sin\Theta_{\rm{T}}\cos\Phi_{\rm{T}}, \sin\Theta_{\rm{T}}\sin\Phi_{\rm{T}}, \cos\Theta_{\rm{T}}\rp,
\end{equation}
where $\Phi_{\rm{L}} = \Phi_E + 20^\circ$ and $\Phi_{\rm{T}} = \Phi_E - 20^\circ$ represent the angular positions of LISA and Taiji relative to Earth’s position $\Phi_E$. The inclination angle $\Theta_{\rm{L/T}} = \pm 60^\circ$ for LISA or Taiji has an undetermined sign.

From the sun's perspective, rotations of the two missions can either align with identical inclination angles or oppose each other with opposite signs of inclination angles~\cite{Joffre:2021hne}. Consequently, the Euler angle $\theta$, parameterizing the relative orientation between LISA and Taiji, remains constant:
\begin{equation}
    \cos\theta =\hat n_{\rm{L}}\cdot\hat n_{\rm{T}}=\cos\Theta_{\rm{L}}\cos\Theta_{\rm{T}} + \cos 40^\circ \sin\Theta_{\rm{L}}\sin\Theta_{\rm{T}}=\frac{1}{4}(1\pm 3\cos40^\circ)\simeq 0.82\text{ or}\,-0.32,
\end{equation}
for aligned and opposite rotations, respectively.

\begin{figure*}[h]
    \centering
    \includegraphics[width=0.55\textwidth]{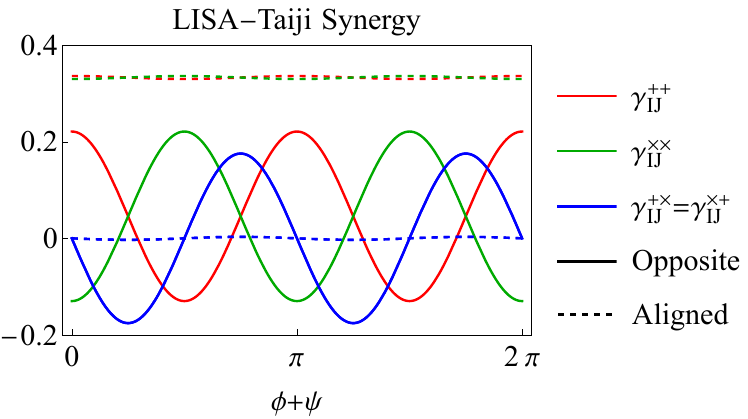}
    \includegraphics[width=0.4\textwidth]{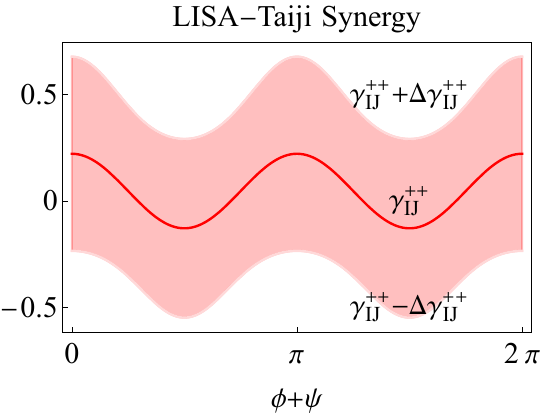}
    \caption{
    {\bf Left:} The four ORFs for the LISA-Taiji pair plotted as functions of the Euler angle combination $\phi+\psi$, for both opposite (solid lines) and aligned (dashed lines) rotations as viewed from the Sun. The static component $\phi - \psi$ is fixed at $0$. The opposite rotation configuration exhibits significantly larger modulations than the aligned configuration.
   {\bf Right:} The ORF $\gamma_{IJ}^{++}$ for the opposite rotation configuration (a subset of the left panel), shown together with its total variance (shaded region) as a function of $\phi+\psi$. The static component $\phi - \psi$ is again fixed at $0$ for simplicity (see Fig.~\ref{fig:config} for the opposite rotation configuration).
}
    \label{fig:LT}
\end{figure*}

The synchronous annual rotation of both missions contributes to a linear increase in the Euler angles $\phi$ and $\psi$ at the same frequency. As a result, the orbital configuration varies as $\phi+\psi$ increases, while $\phi-\psi$ depends on the initial configuration. The orbital average is thus calculated by integrating over $\phi+\psi$ while keeping $\phi-\psi$ constant, defined as: $\langle \cdots \rangle_{\Omega} = \textstyle\int_0^{2\pi}\! \cdots \d(\phi+\psi)/(2\pi)$, resulting in
\begin{align}
&\langle|\tilde{\gamma}^{++}_{\rm{LT}}|^2 \rangle_{\Omega}=\langle|\tilde{\gamma}^{\times\times}_{\rm{LT}}|^2 \rangle_{\Omega}=\frac{1}{2}\cos^8\frac{\theta}{2}\lp 1+\tan^8\frac{\theta}{2}+\cos 4(\phi-\psi)\rp, \label{orbLT1}\\ 
&\langle|\tilde{\gamma}^{+\times}_{\rm{LT}}|^2 \rangle_{\Omega}=\langle|\tilde{\gamma}^{\times+}_{\rm{LT}}|^2 \rangle_{\Omega}=\frac{1}{2}\cos^8\frac{\theta}{2}\lp 1+\tan^8\frac{\theta}{2}-\cos 4(\phi-\psi)\rp.\label{orbLT2}
\end{align}
In the aligned rotation configuration, $\cos^4(\theta/2)$ is significantly larger than $\sin^4(\theta/2)$. This discrepancy results in the variation of ORFs dependent on $\phi+\psi$ being less pronounced compared to the static part controlled by $\phi-\psi$. Conversely, in the opposite rotation configuration, where $\sin^4(\theta/2)$ matches $\cos^4(\theta/2)$ in magnitude. This behavior is clearly illustrated in the left panel of Fig.~\ref{fig:LT}, where the four independent ORFs are plotted for both rotation configurations. We therefore adopt the opposite configuration with $\theta \approx 109^\circ$ for subsequent calculations and, for simplicity, set $\phi - \psi = 0$ initially. The right panel of Fig.~\ref{fig:LT} displays the $\gamma_{IJ}^{++}$ channel for the opposite rotation configuration, with the shaded region indicating the corresponding total variance $\Delta \gamma_{IJ}^{++}$.

For the LISA-Taiji pair, which features comparable sensitivities, the parameter $\xi_I^k$ is approximately uniform across the four channels of both detectors within each frequency bin. In the \emph{short baseline limit}, Eq.~(\ref{eq:TR}) implies that the time required to identify quadrupolar patterns is roughly $5.9$ times longer than that needed for initial detection, based on the orbital average $\langle|\tilde{\gamma}_{IJ}|^2 \rangle_{\Omega} \approx 0.10$. However, when incorporating all relevant frequencies, the most sensitive bins typically correspond to \emph{long baselines} relative to the GW wavelength, which suppresses cross-correlations. As a result, the estimated observation time needed to detect a SGWB power excess is $T_D \approx 0.083$ days, while confirming its quadrupolar nature requires $T_Q \approx 22$ days. Both durations are shorter than the orbital period. Therefore, more precise estimates may vary slightly depending on the actual orbital configuration during this time, since our calculation relies on the orbital-averaged ORFs as an approximation. Note that the observation does not need to be continuous; only each segment must remain uninterrupted.

\subsection{TianQin-AEDGE/MAGIS Synergy}

For correlations between a space-based laser interferometer and an atom interferometer, three channels are available, yielding two possible correlations:
\begin{subequations}\label{c20}
\begin{align}
    \tilde{\gamma}_{\rm T^{+}A}&=\frac{1}{\sqrt{2}}(D^2_{2,0}+D^2_{-2,0})=\frac{\sqrt{3}}{2}\cos2\phi\sin^2\theta, \\
    \tilde{\gamma}_{\rm T^{\times}A}&=\frac{1}{\sqrt{2}i}(D^2_{2,0}-D^2_{-2,0})=\frac{\sqrt{3}}{2}\sin 2\phi\sin^2\theta.
\end{align}
\end{subequations}
The corresponding cross-correlation matrix is then given by:
\begin{equation}
    \tilde{\gamma}=
    \begin{pmatrix}
        0 & 0 & \tilde{\gamma}_{\rm T^+A} \\
        0 & 0 & \tilde{\gamma}_{\rm T^{\times} A} \\
        \tilde{\gamma}_{\rm T^+A}^* &  \tilde{\gamma}_{\rm T^{\times}A}^* & 0 \\
    \end{pmatrix}.
\end{equation}

TianQin consists of three satellites orbiting Earth with an arm length of $L = \sqrt{3} \times 10^5 \,\mathrm{km}$. Its orbital plane is strategically oriented to maintain continuous alignment with the reference binary pulsar source RX J0806.3+1527~\cite{TianQin:2015yph}.

The planned space missions AEDGE and MAGIS will operate in Earth orbit, featuring a single arm that spans  $L = 4.4 \times 10^4 ~ \mathrm{km}$  and separates two satellites by  $138^\circ$  around the Earth~\cite{Graham:2017pmn}. The specific orbital plane for these missions has not yet been determined.

The relative orientation between TianQin and AEDGE/MAGIS is characterized by the separation angle $\Theta$ between their orbital plane normals, along with two rotation angles, $\Phi = \Phi_0 + {2\pi t}/{T_{\rm T}}$ and $\Psi = \Psi_0 + {2\pi t}/{T_{\rm A}}$, which describe the orbital phase evolution of TianQin and AEDGE/MAGIS with respective orbital periods of $T_{\rm T} = 3.15 \times 10^6 \, \mathrm{s}$ and $T_{\rm A} = 3.7 \times 10^5 \, \mathrm{s}$. The Euler angles corresponding to these orbital dynamics are:
\be \theta=\arccos(-\sin\Theta\sin\Psi), \qquad \phi=\Phi+\arctan(\cos\Theta\cot\Psi).\ee
Consequently, the orbital average for this configuration is determined over $\Phi$ and $\Psi$, i.e., $\langle \cdots \rangle_{\Omega} = \textstyle\int_0^{2\pi} \int_0^{2\pi}\! \cdots \d\Phi\d\Psi/(2\pi)^2$, resulting in:
\begin{equation}
\langle|\tilde{\gamma}_{\mathrm{T^{+}A}}|^2\rangle_{\Omega}=\langle|\tilde{\gamma}_{\mathrm{T^{\times}A}}|^2\rangle_{\Omega} =\frac{3}{512}(41+20\cos 2\Theta+3\cos 4\Theta).
\label{eq: TA_orb}
\end{equation}

\begin{figure}[t]
    \centering
    \includegraphics[width=0.5\textwidth]{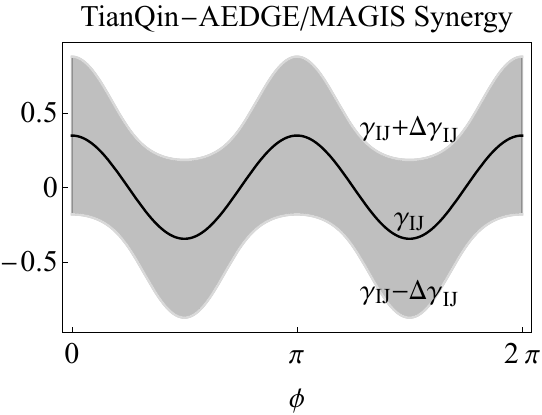}
    \caption{The ORF $\gamma_{IJ} = (\sqrt{3}/5) \cos2\phi$ for the TianQin-AEDGE/MAGIS pair (solid line) and its corresponding total variances (shaded region) are depicted against the varying Euler angle $\phi$ for overlapping orbital planes with $\theta = \pi/2$. The channel for TianQin is selected as $T+$, whereas the ORF involving $T\times$, which is $\propto \sin 2\phi$, differs only by a phase shift of $\pi/2$. Both variables on the respective $x$-axes, $\phi+\psi$ and $\phi$, increase linearly over time.
    }
    \label{fig:LTcorre}
\end{figure}

As the orbit of AEDGE/MAGIS is not yet finalized, we assume $\Theta = 0$, corresponding to overlapping orbital planes as illustrated in the right panel of Fig.~\ref{fig:config}, which maximizes the averaged ORF. In this configuration, $\theta = \pi/2$ remains constant, while $\phi = \pi/2 + \Phi - \Psi$ evolves with a period $T = {T_{\rm A} T_{\rm T}}/({T_{\rm T} \mp T_{\rm A}})$, yielding $4.2 \times 10^5~\mathrm{s}$ or $3.3 \times 10^5~\mathrm{s}$ for aligned and opposite rotations, respectively. The choice of rotation direction does not affect the SNR or quadrupolar identification, as both yield the same orbit-averaged ORF. In this setup, the ORF involving the $T+$ channel of TianQin is given by $\gamma_{IJ} = (\sqrt{3}/5)\cos 2\phi$, as shown in Fig.~\ref{fig:LTcorre}, while the ORF for the $T\times$ channel differs only by a $\pi/2$ phase shift.

For the TianQin-AEDGE/MAGIS pair, the required durations extend to approximately $T_D\approx3.2$ days for initial detection and $T_Q\approx42$ days for quadrupolar identification. These differences primarily arise from distinct frequency sensitivities and orbital considerations. 


\section{Discussion}\label{sec:discussion}
Distinguishing the SGWB from noise requires characterizing its quadrupolar nature, observable through cross-correlations between detector pairs. Using several space missions as illustrative examples, we classify detector channels based on their antenna responses and construct the corresponding ORFs, which depend on the Euler angles relating the detector frames. We demonstrate how orbital and orientational motions of these detectors effectively scan ORFs across configuration space via Euler angles, facilitating the identification of the SGWB’s quadrupolar structure. This approach highlights the enhanced observational capabilities achievable through GW detector networks~\cite{Klimenko:2005xv,Ruan:2019tje,Ruan:2020smc,Zhang:2020hyx,Wang:2020dkc,Orlando:2020oko,Chen:2021sco,Wang:2021mou,Wang:2021srv,Wang:2021uih,Wang:2021njt,Zhang:2021wwd,Cai:2023ywp,Jin:2023sfc,Liang:2023fdf,Li:2023szq,Mentasti:2023uyi,Zhao:2024yau,Marriott-Best:2024anh}.

We find that the duration necessary for quadrupolar identification exceeds by approximately one to two orders of magnitude the time required for the initial detection of a power excess. Specifically, for the LISA-Taiji pair, the optimal configuration involves opposite rotations, maximizing variations in their relative orientation according to their current orbital designs, thereby significantly improving quadrupolar pattern detection.

Our methodology extends beyond space-based missions, encompassing broader GW detector networks such as tabletop high-frequency detectors~\cite{Li:2003tv,Arvanitaki:2012cn,Berlin:2021txa,Domcke:2022rgu,Berlin:2023grv,Gao:2023gph,Schmieden:2023fzn,Alesini:2023qed,Chen:2023ryb,Gao:2023ggo,Kahn:2023mrj,Navarro:2023eii,Gatti:2024mde,Valero:2024ncz,Domcke:2024mfu,Schneemann:2024qli,Domcke:2024eti}, astrophysical observations~\cite{Hui:2012yp,Blas:2021mpc,Blas:2021mqw,Fedderke:2021kuy,Fedderke:2022kxq,DeRocco:2023qae,Alves:2024ulc,Zwick:2024hag,Lu:2024yuo,Crosta:2024udx,Wang:2024tnk}, and lunar-based detectors~\cite{Yan:2024jio,Ajith:2024mie,Yan:2024jik}, each characterized by unique response functions. Integrating these diverse systems into a cohesive network with optimized orientations and orbital configurations promises to substantially enrich the information extracted from future GW observations.

\section*{Acknowledgements} 
We are grateful to Huaike Guo, Ziren Luo, Gang Wang, and Yue Zhao for useful discussions.
This work is supported by the National Key Research and Development Program of China under Grant No. 2020YFC2201501. 
Y.C. is supported by VILLUM FONDEN (grant no. 37766), by the Danish Research Foundation, and under the European Union’s H2020 ERC Advanced Grant “Black holes: gravitational engines of discovery” grant agreement no. Gravitas–101052587, and by FCT (Fundação para a Ciência e Tecnologia I.P, Portugal) under project No. 2022.01324.PTDC.
J.S. is supported by Peking University under startup Grant No. 7101302974 and the National Natural Science Foundation of China under Grants No. 12025507, No.12150015; and is supported by the Key Research Program of Frontier Science of the Chinese Academy of Sciences (CAS) under Grants No. ZDBS-LY-7003.
IFAE is partially funded by the CERCA program of the Generalitat de Catalunya. X.X. is funded by the grant CNS2023-143767. 
Grant CNS2023-143767 funded by MICIU/AEI/10.13039/501100011033 and by European Union NextGenerationEU/PRTR.
Y.C. and X.X. acknowledge the support of the Rosenfeld foundation and the European Consortium for Astroparticle Theory in the form of an Exchange Travel Grant.


\end{document}